\documentclass[twocolumn,showpacs,preprintnumbers,amsmath,amssymb,prb]{revtex4}

\usepackage{graphicx}
\usepackage{dcolumn}
\usepackage{bm}
\usepackage{wasysym}
\usepackage{subfigure}
\usepackage{float}
\usepackage{wrapfig}

\begin{document}

\preprint{}

\title{Half-metallicity induced by charge injection in hexagonal-BN clusters embedded in graphene}

\author{Marcos G. Menezes$^1$}
 \email{marcosgm@if.ufrj.br}
\author{Rodrigo B. Capaz$^{1,2}$}%
 \email{capaz@if.ufrj.br}
\affiliation{$^1$ Instituto de F\'{i}sica, Universidade Federal do Rio de Janeiro, Caixa Postal 68528 21941-972, Rio de Janeiro, RJ, Brazil \\ $^2$ Divis\~ao de Metrologia de Materiais, Instituto Nacional de Metrologia, Normaliza\c c\~ao e Qualidade Industrial $-$ Inmetro, Duque de Caxias, Rio de Janeiro, 25245-020, Brazil}%

\date{\today}

\begin{abstract}
\b
We study the electronic structure and magnetic properties of h-BN triangular clusters embedded in graphene supercells. We find that, depending on the sizes of the clusters and the graphene separation region between them, spin polarization can be induced through charge doping or can be observed even in the neutral state. For these cases, half-metallicity is observed for certain charged states, which are otherwhise mettalic. In these half-metallic states, the spin density is concentrated near the edges of the clusters, in analogy to the more common predictions for half-metals in zigzag graphene nanoribbons and h-BN/graphene intercalated nanoribbons. Since experimental realizations of h-BN domains in graphene have already been reported, these heterostructures can be suitable candidates for nanoelectronics and spintronics applications.
\end{abstract}

\pacs{71.15.Mb,73.22.Pr,75.70.Cn}
\maketitle

\section{\label{intro}Introduction}
Since its experimental observation \cite{novoselov_science}, graphene, a two-dimensional carbon sheet with a honeycomb structure, has attracted a lot of attention from the scientific community  due to its unique properties, such as massless Dirac fermions, anomalous quantum hall effect and many others, which lead to several potential applications in nanoelectronics \cite{neto_rmp}. This excitement with graphene led to further research on systems sharing the same crystal structure, such as the h-BN sheet, which has also been observed experimentally \cite{watanabe_natmat,kubota_science}, and other group III-V systems \cite{wu_applphyslett}. In contrast to graphene, which is a zero-gap semiconductor, h-BN is a wide-gap insulator, with a band gap value of up to $5.9$ eV \cite{watanabe_natmat,kubota_science}. This suggests that a possible route to use graphene in a semiconductor device with a tunable gap, a very desirable property in nanoelectronics, would be to use hybrid h-BN/graphene sheets, in which the gap could be tuned by the relative size of the h-BN and graphene regions, given that the resulting heterostructure is formed by h-BN and graphene domains instead of an alloy of BNC. Such domain structures have already been observed experimentally, and they show properties midway between those of pure h-BN and pure graphene, thus providing support for this route \cite{ci_natmat}. Recently, an anomalous metal-to-insulator transition
has been observed in this system\cite{rodrigo}. In the field of spintronics, another desirable property for materials is half-metallicity, which consists of a metallic behavior for one spin channel and a semiconducting one for the other. Some half-metals have already been predicted in graphene and h-BN related one-dimensional structures, such as graphene zigzag ribbons with an in-plane external electrical field \cite{son_nature} and h-BN/graphene intercalated ribbons \cite{pruneda_prb,bhowmick_jphyschem}. Half-metallicity has also been predicted for the pure h-BN sheet with electron doping \cite{wu_applphyslett}.

In this work, we study the electronic and magnetic structure of h-BN triangular clusters embedded in graphene. We predict that half-metallicity can be induced through charge injection in this system, depending on the sizes of the clusters and the graphene separation region. By analyzing the band structure and the spin density dependence on the sizes, we find that spin polarization is observed whenever the Fermi level is close to weakly-dispersive states, with their bandwidths and mean kinetic energies below $\sim 0.1$ eV . When spin polarization is observed, we have either metallic or half-metallic states depending on the injected charge. These states are characterized by a concentration of the spin density on the carbon atoms near the edges of each cluster, a feature already observed in antidot lattices, single h-BN clusters in graphene and other graphene or h-BN related nanostructures \cite{yazyev_prl,xu_japp,yazyev_review}. The properties of both localized and non-localized bands depends on the relative sizes of the regions, giving rise to three types of band structure, and consequently magnetic properties, which we will discuss later. Therefore, we can tailor the electronic and magnetic properties of this system by controlling the sizes of the clusters, the graphene region and doping, which makes this system a suitable candidate for nanoelectronics and spintronics applications.

Our paper is organized as follows. In the next Section, we present our theoretical methodology. In Section \ref{bstruct}, we discuss the band structure results and electronic properties of the different studied configurations and in Section \ref{mag}, we analyze the magnetic properties, by discussing the spin density and projected density of states results. Finally, in Section \ref{conclusion}, we present our conclusions and provide a discussion of how the effects observed in our model systems would be expected in a more general and realistic situation, that is, for non-periodic systems and h-BN domains of arbitrary shapes, as observed in experimental grounds.

\section{\label{met}Methodology}

In this work, we employ self-consistent calculations within density functional theory (DFT) \cite{ho-kohn,kohn-sham}, as implemented in the SIESTA code \cite{siesta}$^{,}$ \footnote{Within SIESTA, we use norm-conserving Troullier-Martins pseudopotentials \cite{troullier-martins} for the ion-electron interaction, the PBE-GGA approximation for the exchange-correlation functional \cite{pbe-gga} and a single-zeta (SZ) pseudoatomic basis set for the expansion of the electronic wavefunctions. Even with a minimum basis, SIESTA should have a higher accuracy than that of semi-empirical calculations based on tight-binding and mean-field Hubbard model, which have been used extensively and successfully to study magnetic properties of graphene-like systems \cite{yazyev_review,yazyev_prl,soriano_prl}. The use of a double-zeta-polarized (DZP) basis set on selected test configurations doesn't give much improved results over the SZ basis. The real space grid energy cutoff is set to $150$ Ry. We also use a $6 \times 6 \times 1$ Monkhorst-Pack k-point sampling \cite{monkhorst-pack} and an electronic smearing (Fermi-Dirac type) of $100.0$ K in all calculations.}. For all neutral systems studied, we let the atomic coordinates relax in all directions until all the forces are smaller than $0.04$ eV\slash\AA \ . No significant deviations from the planar position (buckling) were observed after the relaxation. For the charged states, we use the same relaxed coordinates of the corresponding neutral state. The supercell lattice constant is set to $N_s a$, where $N_s$ is the supercell dimension and $a = 2.47$ \AA \ is the pure graphene lattice constant. Since graphene and h-BN have similar lattice constants, we only let the atomic coordinates relax and the supercell volume is fixed. All supercells have a $10$ \AA \ spacing in the \textit{z} direction in order to eliminate spurious interactions between periodic images. Even tough there are local in-plane dipole moments present in the supercell (pointing from one cluster to its neighbors), no stray electrical fields are present in the vacuum region.

\begin{figure}[h]
\centering
\includegraphics[width=8cm]{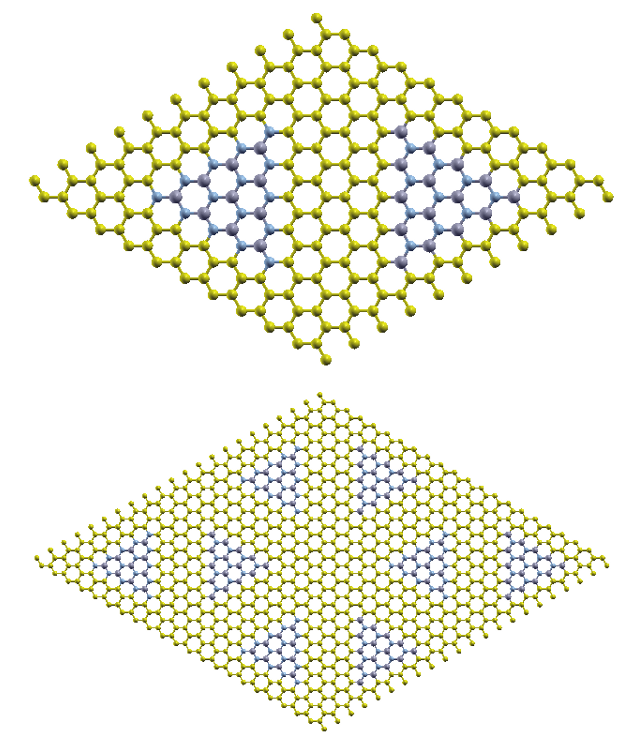}
\caption{\label{cell} Top: A $11 \times 11$ supercell containing two triangular $B_{15}N_{10}$ and $B_{10}N_{15}$ ($\Delta = 5$) clusters, being a representative of all cells used in this work. The triangular clusters are facing opposite directions, with two parallel edges. Notice that one cluster contains only B atoms on the edges, whereas the other contains only N atoms. Bottom: The same supercell pictured with 3 periodical images, showing the regular spacing of the clusters by $N_z = 4$ graphene zigzag chains in all directions.}
\end{figure}

A representative $11 \times 11$ supercell containing two triangular $B_{15}N_{10}$ and  $B_{10}N_{15}$ clusters is shown in the top panel of figure \ref{cell}. All the supercells considered in this work share the same geometrical features: the two triangular clusters always have the same size and face opposite directions, with two parallel edges. The edges of one cluster are terminated by B atoms, whereas the edges of the other are terminated by N atoms. The spacing of the clusters, characterized by the number $N_z = 4$ of graphene zigzag chains between them, is the same in all directions, as shown in the bottom panel of the same figure, which pictures four periodic images of the supercell. Notice that $N_z$ is not an independent parameter, because it is fixed by the clusters' size and the dimension of the supercell. Therefore, in a general supercell, the main defining parameters are $N_s$ (or $N_z$) and the size of the clusters. The latter is characterized by the parameter $\Delta = |m - n|$ for a triangular cluster with general stoichiometry $B_mN_n$. It can be shown that the number of atoms in the edge of the cluster (B or N) is $\Delta$ and the total number of atoms $m + n$ within one single cluster is $\Delta^2$. The clusters' sizes studied in this work range from $\Delta = 2$ ($B_3N_1$ and $B_1N_3$) to $\Delta = 9$ ($B_{45}N_{36}$ and $B_{36}N_{45}$), and the supercell sizes range from $8 \times 8$ to $ 15 \times 15$. Not all combinations of clusters and supercells give an uniform spacing in all directions, so one must be careful with this choice. This is shown in detail in Table \ref{tab_studied}, where we show all the configurations we studied. The diagonals of this table represent systems with clusters of different sizes, but the same spacing $N_z$. Note also that odd values of $N_z$ aren't allowed for this system, since it's impossible to construct two clusters of the same size or shape. We will analyze our results mainly in terms of $N_z$ and $\Delta$. The relation between these two parameters and $N_s$ is given by $N_s = \Delta + 3N_z/2$.

The reason for the choice of the triangular shape for the clusters and their arrangement in this work is that previous {\it ab-initio} calculations of single hexagonal and triangular clusters embedded in graphene supercells shows that, while the former in nonmagnetic, the latter is ferromagnetic, with opposite spin alignment in the A and B sublattices \cite{xu_japp}. Test calculations we performed on single clusters do reproduce this result. Since all the edges of a single triangular cluster lie only on one sublattice and, with our choice of supercell geometry, the edges of different neighboring clusters lie on different sublattices, we expect to obtain an antiferromagnetic coupling between the clusters. As we shall see, this is indeed what we obtain for the large clusters and graphene regions in neutral configurations. On the other hand, the smaller systems are found to be in a delocalized, weak antiferromagnetic state (essentialy non-magnetic), which seems to be due to a cancelation effect between neighboring edges. Moreover, since the parallel edges of neighboring clusters have different kinds of atoms (B or N), we will have an electrical field in the graphene region between them, as a result of the polarization of the system. This kind of arrangement is similar to the one used in the study of h-BN/graphene intercalated zigzag ribbons \cite{pruneda_prb,bhowmick_jphyschem}, where a half-metal was found.

\begin{table}
\scalebox{0.85}{
\begin{tabular}{ c | *{8}{c} }
       & $B_3N_1$ & $B_6N_3$ & $B_{10}N_6$ & $B_{15}N_{10}$ & $B_{21}N_{15}$ & $B_{28}N_{21}$ & $B_{36}N_{28}$ & $B_{45}N_{36}$ \\
$\Delta$ & 2      & 3      & 4       & 5        & 6        & 7        & 8        & 9 \\
\hline
$N_s$  \\
8      & 4*     & -      & -       & 2*       & -        & -        & -        &   \\
9      & -      & 4*     & -       & -        & 2        & -        & -        & - \\
10     & -      & -      & 4*      & -        & -        & 2        & -        & - \\
11     & 6*     & -      & -       & 4*       & -        & -        & 2*       & - \\
12     & -      & 6*     & -       & -        & 4*       & -        & -        & 2*\\
13     & -      & -      & 6       & -        & -        & 4*       & -        & - \\
14     & 8*     & -      & -       & 6*       & -        & -        & 4*       & - \\
15     & -      & 8*     & -       & -        & 6*       & -        & -        & 4*\\
\end{tabular}
}
\caption{\label{tab_studied} Possible configurations for the supercell geometry considered in this work. The entries in the table are the values of $N_z$ for the given configuration and the values marked with a star represent the configurations we studied. The selected configurations spam a wide range of values for $N_z$ and $\Delta$, allowing us to obtain the general trends for this system.}
\end{table}

\section{\label{bstruct} Band Structure Results}

\begin{figure*}
\centering
\includegraphics[width=16cm]{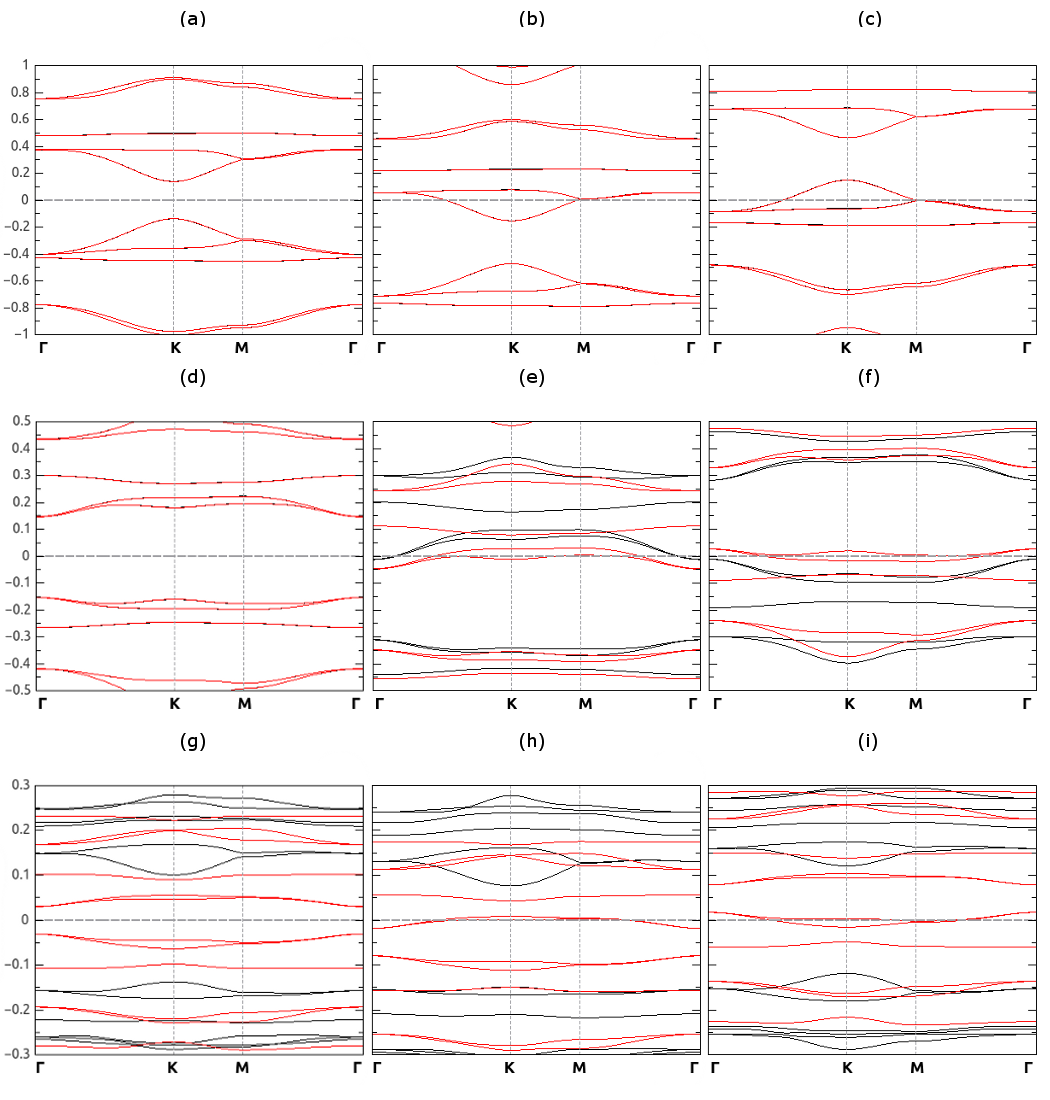}
\caption{\label{bands} Selected results of the three general types of band structures observed in this system, as described in the text. These results are for a constant $N_z = 4$ graphene separation, but different values of $N_z$ show similar trends. Energies are in eV and the Fermi level is set to zero in all panels. Note that the energy scale is different in each row. Spin up (down) bands are represented by black (red) lines. (a)-(c): type 1 - $B_6N_3$ + $B_3N_6$ ($\Delta=3$) and $N_s=9$. (d)-(f): type 2 - $B_{15}N_{10}$ + $B_{10}N_{15}$ ($\Delta=5$) and $N_s=11$. (g)-(i): type 3 - $B_{45}N_{36}$ + $B_{36}N_{45}$ ($\Delta=9$) and $N_s=15$. All the bands on the left column are for neutral states, while the middle column are for -1 charged states and the right column for $+1$ charged states. Note that the $+1$ type 2 and $\pm 1$ type 3 charged states are half-metallic.}
\end{figure*}

In our calculations, we observe three general types of band structure in this system, which we will call type 1, 2 and 3 throughout this paper. In Fig. \ref{bands}, we show examples of each of these types for neutral (left column), -1 (middle column) and +1 (right column) charged states for a constant $N_z = 4$ graphene separation. In the type 1 (top row) band structure, observed for lower cluster sizes ($\Delta = 2 - 4$), we see two pairs of dispersive bands, followed by a pair of weakly-dispersive states near the neutral Fermi level. These bands have no spin polarization in the neutral state and remain unpolarized as we populate (unpopulate) the dispersive bands through injection of an electron (hole), as shown in panels (a)-(c) In principle, if we keep doping this system the Fermi level will eventually cross the weak, almost non-dispersive levels. Therefore, we expect spin polarization in this case, due to the usual zero energy instability effect, in which the zero kinetic energy of the level allows the splitting of spin up and down bands due to the exchange interaction, thus removing the instability. However, this possibility would correspond to extremely high doping levels and therefore it wasn't tested in this work.

For intermediary cluster sizes ($\Delta = 5 - 7$), we observe the type 2 band structure, depicted in the middle row of Fig. \ref{bands}. In this case, we can see the same set of dispersive and non-dispersive bands as in the previous case, but the bandwidth of the former is lower and both are closer to the neutral Fermi level. These bands still remain unpolarized in the neutral state, but now they polarize in the charged states, as shown in panels (d)-(f). This happens because now the bandwidth of the dispersive bands is lower than a limit value that determines the zero energy instability, such that the mean kinetic energy of these bands when the Fermi level crosses them is close to zero. Moreover, for $\Delta \geq 6$, we observe that another pair of weakly-dispersive bands appears close to the neutral Fermi level, which now is the closest pair. These levels also show spin polarization when crossed by the Fermi level (thus also being type 2), providing positive evidence for a similar behavior of these levels in type 1 bands, as described in the previous paragraph.

Finally, for the largest cluster sizes studied in the $N_z = 4$ family ($\Delta \geq 8$), we observe the type 3 band structure, which is shown in the bottom row of Fig. \ref{bands}. In this case, all the bands near the neutral Fermi level have a very small bandwidth (smaller than $0.05$ eV), which already favors spin polarization in charged states, as in type 2. Moreover, we have a new set of weakly-dispersive bands even closer to the Fermi level. As we can see in the left panel, this allows for spin polarization also in the neutral state, where the mean energy of the closest spin down bands is smaller than $0.05$ eV. By combining our results from the three types of band structure observed, we estimate that the maximum bandwidth that a band crossed by the Fermi level must have in order to observe spin polarization in the charged configuration is around $0.1$ eV. If the mean kinetic energy of the closest band is also of this same order, we also expect spin polarization to occur in the neutral state. This should be the order of magnitude for the exchange interaction in this system, which is responsible for the splitting of the bands.

\begin{figure*}
\centering
\includegraphics[width=16cm]{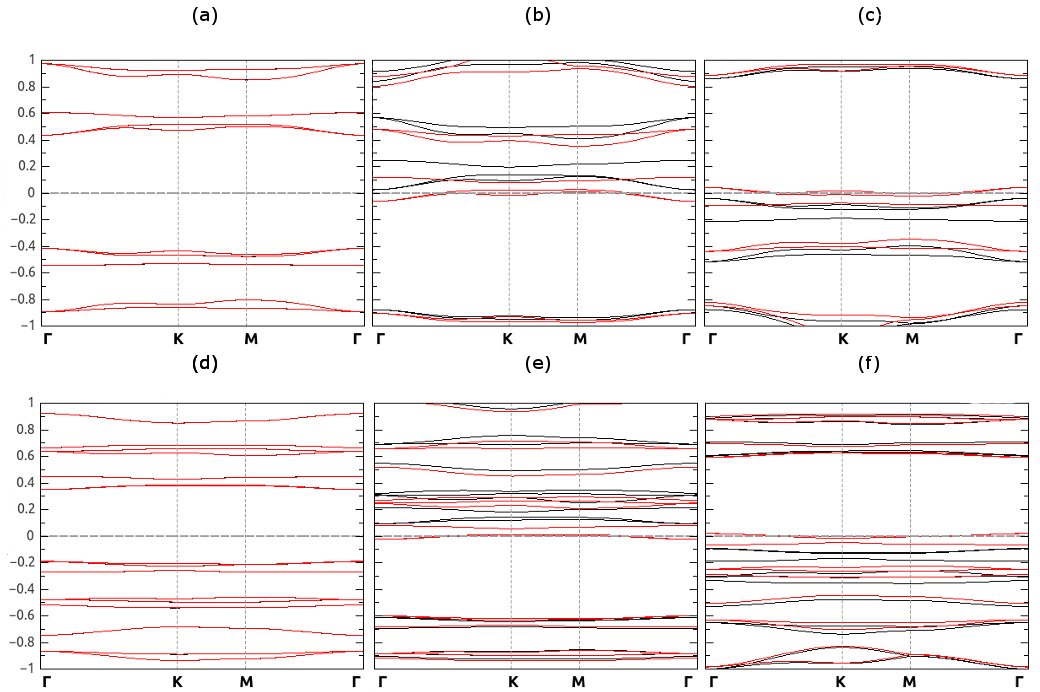}
\caption{\label{bands2} Selected results for the $N_z = 2$ family. The units, lines and column organization are the same as in Fig. \ref{bands}. (a) - (c): $B_{15}N_{10}$ + $B_{10}N_{15}$ ($\Delta = 5$) and $N_s = 8$. (d)-(f): $B_{45}N_{36}$ + $B_{36}N_{45}$ ($\Delta = 9$) and $N_s = 12$. We have only observed the type 2 band structure in this family in the range of values of $\Delta$ considered in this work. Nevertheless, since the band gap decreases with increasing $\Delta$, we expect to see a type 3 behavior for larger cluster sizes. Note also that all the charged states shown in this picture are half-metallic.}
\end{figure*}

So far we have analyzed only the electronic structure behavior of our system as a function of $\Delta$ for a fixed graphene separation $N_z = 4$, because this family spans the largest number of configurations and varieties of band structure we studied. In Table \ref{types}, we show our results for all configurations considered in this work. We see that similar trends are observed in other families, but not all the three types are observed. In Fig. \ref{bands2}, we show our results for the $N_z = 2$ family, using the same cluster sizes as in panels (d)-(f) of Fig. \ref{bands}. We see that the bandwidths of the closest bands are roughly the same as in the respective cases of the $N_z = 4$, but their mean kinetic energy is higher, thus making spin polarization in the neutral state more difficult. This explains why the type 2 band structure is observed through a wider range of cluster sizes in this family. In general, for a fixed value of $\Delta$, as we increase the value of $N_z$ these bands tend to get closer to the neutral Fermi level, so the band gap decreases as well, as expected from the pure graphene supercell limit $\Delta = 0$. Since this size dependence is similar to that of graphene nanoribbons, we can think of our system as a series of interconected zigzag graphene nanoribbons separated by the h-BN clusters, as can be seen from the periodic images of our supercell shown in the bottom panel of Fig. \ref{cell}. Therefore, this structure is analogous to the h-BN/graphene ribbon superlattices already studied in the literature \cite{pruneda_prb,bhowmick_jphyschem,liu_jphyschem}. In contrast, the bandwidth of the closest bands doesn't seem to depend on $N_z$ in the range considered, but it decreases with increasing $\Delta$, which, thus indicating increasing localization of the charge carriers. As we shall see in the next section, such localization behavior is also observed for the spin density, with a concentration on the C atoms near the edges of the clusters.

\begin{table}
\scalebox{0.85}{
\begin{tabular}{ c | *{8}{c} }
       & $B_3N_1$ & $B_6N_3$ & $B_{10}N_6$ & $B_{15}N_{10}$ & $B_{21}N_{15}$ & $B_{28}N_{21}$ & $B_{36}N_{28}$ & $B_{45}N_{36}$ \\
$\Delta$ & 2 & 3 & 4 & 5 & 6 & 7 & 8 & 9 \\
\hline
$N_s$  \\
8      & 1 & - & - & 2 ($\pm 1$) & - & - & - &   \\
9      & - & 1 & - & - & - & - & - & - \\
10     & - & - & 1 & - & - & - & - & - \\
11     & 1 & - & - & 2 ($+1$) & - & - & 2 ($\pm 1$) & - \\
12     & - & 1 & - & - & 2 ($\pm 1$) & - & - & 2 ($\pm 1$) \\
13     & - & - & - & - & - & 2 ($\pm 1$) & - & - \\
14     & 1 & - & - & 2 & - & - & 3 ($-1,\pm 2$) & - \\
15     & - & 1 & - & - & 2 ($-1$) & - & - & 3($\pm 1,+2$) \\
\end{tabular}
}
\caption{\label{types} Types of band structure observed for each configuration we tested. For a given value of $N_z$ (one diagonal) and increasing cluster size $\Delta$, the band structure changes its character from type 1, with no spin polarization neither in neutral or $\pm 1$ charged states, to type 2, with no spin polarization in the neutral state, but polarized for charged states and finally to type 3, with spin polarization in neutral and charged states. This is related to the position of the weakly-dispersive levels as a function of $\Delta$ and $N_z$, as observed in Fig. \ref{bands}. Half-metallic states were observed for some type 2 and 3 charged states, as shown in parenthesis in the entries of the table.}
\end{table}

To complete our dicussion on the band structure results, we now consider the electronic properties. First, note that, in our supercell, the C atoms bonded to the B atoms on the edge of one cluster and those bonded to the N atoms on the edge of the other belong to different sublattices of the honeycomb structure. This breaks the inversion symmetry of the lattice, thus rendering the A and B sites inequivalent and opening a band gap in the pure graphene band structure. Therefore, all possible configurations for this supercell should be semiconducting in the neutral state, which is indeed observed for all configurations considered here. Moreover, since the graphene region between the clusters is conected to B atoms on one side and N atoms on the other, a polarization field is induced in this region, in a similar fashion to the analogue discussed previously. In this case, a half-semimetallic (semiconductor for one spin channel and a very small gap for the other) was observed in the neutral state, and possibly a half-metal could be obtained in charged states, a possibility not tested in that work. In our case, we do observe half-metallic states in type 2 and type 3 bands for many charged configurations. A few examples can be seen in the middle and right columns of Figs. \ref{bands} and \ref{bands2} and all the cases are listed in parenthesis in the entries of Table \ref{types}. We studied $\pm 1$ charged states in all calculations, and also $\pm 2$ on a few ones, which corresponds to an injected charge density range between $10^{12}$ and $10^{13}$ cm$^{-2}$ for the supercell sizes considered. The charged states not shown in the table were either not considered or are regular metals. As we shall see in the next section, in all charged configurations (whether metallic or half-metallic), the charge carriers are typically localized on the carbon atoms close to the edges of one of the clusters, depending of their sign (electrons or holes).

\section{\label{mag} Magnetic Properties and Density of States}

\begin{figure*}
\centering
\includegraphics[width=16cm]{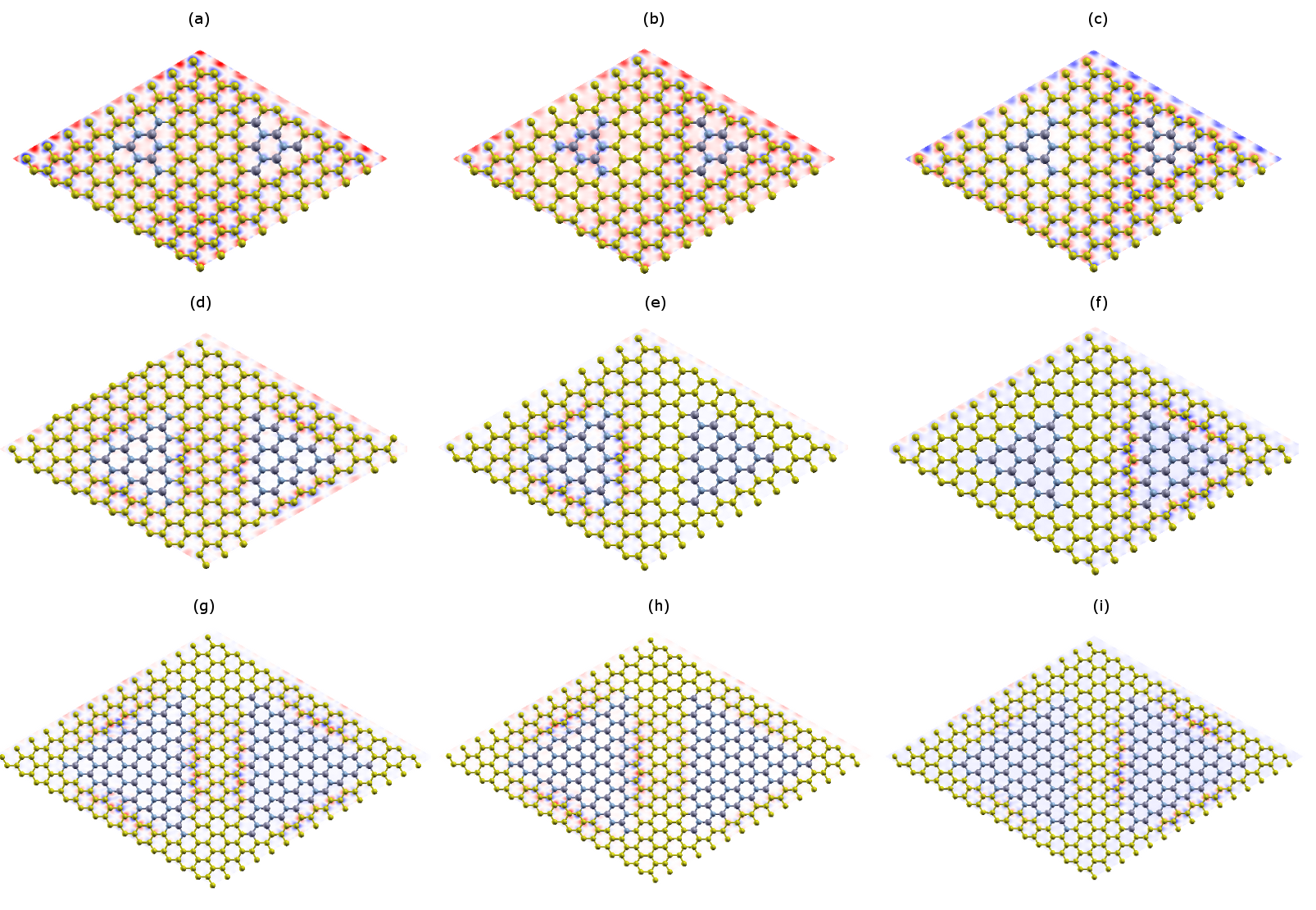}
\caption{\label{spindns} Spin density results for the $N_z = 4$ family. The selected configurations are the same as in Fig. \ref{bands}. Spin up values are shown in red and spin down in blue. Maximum absolute value for the spin density is $\approx 10^{-5}$ bohr$^{-3}$ for type 1 (a)-(c) and neutral type 2 (d), and $5 - 10 \times 10^{-4}$ bohr$^{-3}$ for charged type 2 (e)-(f) and type 3 (g)-(i). Note the increasing intensity and degree of localization of the spin density on the carbon atoms near the edges of each cluster with increasing cluster size, with electrons tending to localize near the N terminated cluster and holes around the B terminated one.}
\end{figure*}

As discussed in the previous section, the C atoms bonded to the N and B atoms on the edges of each cluster belong to different sublattices in our system. Therefore, since in pure graphene and other graphene related systems the C atoms in the A and B sublattices tend to have opposite spin alignment when the system is spin polarized, we expect to see ferromagnetic (FM) coupling between the edges of one single cluster and antiferromagnetic (AF) coupling between the edges of neighboring clusters in the neutral state. This is indeed what we observe for type 2 and type 3 configurations, as shown in the spin density profiles in Fig. \ref{spindns} for our standard examples in the $N_z = 4$ family. For type 1, we have an overall AF state with a very small spin polarization (maximum absolute value smaller than $10^{-5}$ bohr$^{-3}$), which resembles the pure graphene AF state and can be regarded essentially as a non-magnetic (NM) state. As we increase the cluster size and reach type 2 configurations, we see that the neutral state already shows localization of the spin density in the C atoms near the edges of the clusters and in the graphene separation region. However, the maximum absolute value of the spin density in this case is of the same order as in type 1. This situation changes when we inject charge in the system, which induces spin polarization in the band structure and increases the spin density amplitude about 50 times. In type 3 configurations, the degree of localization near the edges gets even higher, with a profile (in the graphene separation region) very similar to the one observed in zigzag graphene nanoribbons and in the 1D intercalated h-BN/graphene ribbons analogue of our system discussed in the last section. In this case, the neutral state is already spin polarized and the spin density amplitude is around $10^{-3}$ bohr$^{-3}$, which is 100 times larger than the values for type 1 and neutral type 2. The charged states have amplitudes of the same order of magnitude.

\begin{figure}
\centering
\includegraphics[width=8cm]{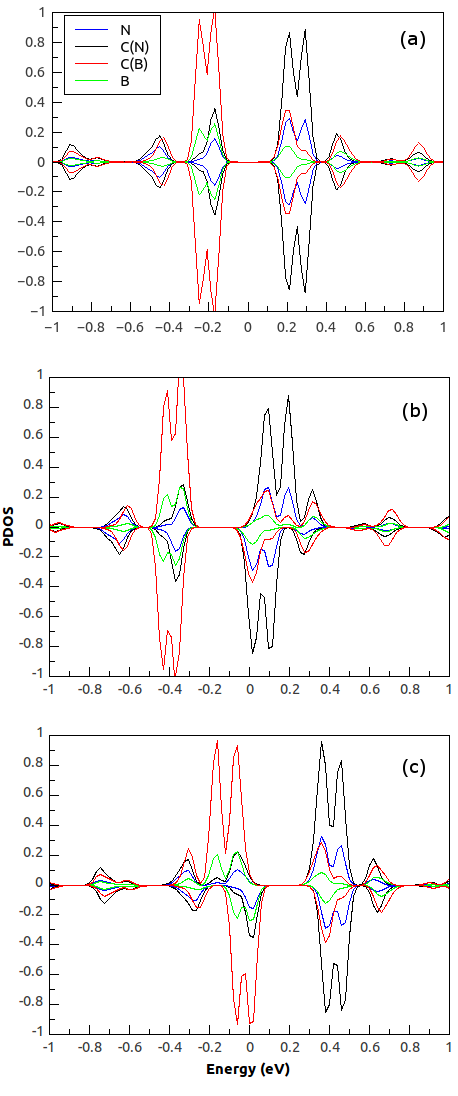}
\caption{\label{pdos} Projected density of states for $\pi_z$ orbitals of selected atoms in the $B_{15}N{10}$ + $B_{10}N_{15}$, $N_s = 11$, $N_z = 4$ configuration (type 2). (a) Neutral state; (b) -1 charged state; (c) +1 charged state. The color codes for the different curves is shown in panel (a) and the notation is described as follows: B (N): Boron (Nitrogen) atom on the edge of one cluster; C(B) (C(N)): Carbon atom bonded to the B (N) atom on the edge. For the edge atoms we choose those on the central region of the edge, because spin polarization is reduced near the vertexes. As always, the Fermi level is set to zero in all cases.}
\end{figure}

We can also note from Fig. \ref{spindns} that, for types 2 and 3, the charged states have a stronger spin density magnitude localized near the N terminated cluster for electron doping and near the B terminated cluster for hole doping. To understand this effect, we need to analyze the nature of the occupied and unoccupied bands near the neutral Fermi level, so, we performed projected density of states calculations (PDOS) for the B, N and C atoms that define the edges of the clusters. We choose atoms from the middle of the edges, because the atoms near the vertices show a smaller spin polarization amplitude. In Fig. \ref{pdos}, we show our results for the type 2 example in the $N_z = 4$ family, and similar trends are observed for other types and families (of course, the differences between the PDOS of the three types will reflect those of their band structures). We only show the PDOS of $p_z$ orbitals in the picture, since the $\sigma$ orbitals are found to have no significant contribution on the energy scale of the picture. The C atom bonded to the N atom on the edge of one of the clusters (labeled C(N), black lines) gives its greatest contribution to the energy levels right above the Fermi level (between $0.1 - 0.35$ eV, compare with Fig. \ref{bands}, panels (d-f)). The N atom PDOS (blue lines) follows the same behavior, but with a magnitude smaller than the C(N) atom curve, indicating that the energy levels in this range are localized states, mainly formed by the hybridization of $p_z$ orbitals of C(N) and N edge atoms. Similarly, the C(B) and B edge atoms of the other cluster (red and green lines, respectively) give their main contributions to the levels right below the neutral Fermi level, which means these bands are also localized states, mainly formed by hybridization of $p_z$ orbitals from these atoms. We also point out that PDOS calculations for the C atoms on the center of the graphene separation region (not shown in the picture) show that these atoms give much smaller contributions to the energy levels discussed above, having the same order of magnitude as the average PDOS per atom in this range, which provides more evidence for these levels being localized states.

When we inject one electron in the system (panel b), we see a relative shift in energy between the spin up and spin down PDOS, revealing the spin polarization observed in type 2 configurations. This shift is much greater for the energy levels closest to the neutral Fermi level, which are the levels whose main contributions come from the C(N) and N edge atoms, as discussed above. On the other hand, the lower levels, that is, those related to the C(B) and B edge atoms, have a small shift and therefore a small spin density. This explains the higher magnitude of the spin density near the N terminated cluster in this case, as seen in Fig. \ref{spindns}. Similarly, for hole doping (panel c), the Fermi level is closest to the C(B) and B energy levels, so now these levels have the largest shift and the spin density amplitude is higher near the B terminated clusters. For type 1 configurations, we don't see any shift between the spin up and spin down PDOS for any set of levels, which agrees with the observed band structure. The spin density resembles more that of the pure graphene sheet in the AF state, altough the spin density seems stronger on the B terminated cluster for charged cases. In the case of type 3, both set of levels are close to the neutral Fermi level, so we see a shift in the PDOS for both of them, which explains the spin density concentration near both clusters and also agrees with the band structure. Finally, for the atoms of the triangular clusters in types 2 and 3, the proximity of two consecutive edges seems to lead to a cancelation effect for the spin density. The levels coming from the two edges hybridize and are pulled further away from the neutral Fermi level, thus inhibiting spin polarization. This could also explain the observed band structure and spin density for smaller clusters, which are type 1 (note that the spin density scale is much smaller in panels a-d of Fig. \ref{spindns} than in panels in e-i).

To finish this section, we now discuss the macroscopic magnetization of the supercell. In the literature, it is common to interpret the results using Lieb's theorem, which states that, for a bipartite lattice and antidot defects, the net magnetization is proportional to $N_A - N_B$, where $N_A$ ($N_B$) is the number of atoms missing from the A (B) sublattice \cite{lieb_prb}. In our system, we can think of the h-BN clusters as antidots deffects because, as seen from Fig. \ref{spindns}, the spin density doesn't penetrate much into the clusters and the energy levels associated with them lie further away from the Fermi level. In this sense, the levels we have discussed so far to explain the electronic and magnetic properties observed can be considered "midgap states". However, as we have already shown, these levels consist mainly of a hybridization of C and B or N atoms on the edges and they show some dispersion, in contrast with pure antidots, which only have C atoms and less dispersive levels \cite{neto_rmp,vozmediano_prb}. This leads to some important deviations from Lieb's theorem, which were already observed in single cluster calculations \cite{xu_japp}. Nevertheless, in our system we find that all neutral cases show no net magnetization, in agreement with the theorem. It predicts $M = 0$ because, in our supercell, all B atoms lie in one sublattice, whereas all N atoms lie on the other, thus $N_A = N_B$. On the other hand, for the charged states, we observe non zero values for the magnetization in types 2 and 3 and it is specially strong for the half-metals, where $M = q \mu_B$ and $q$ is the value of the injected charge in units of $e$ (this corresponds to a full polarization of the injected charge). If the additional charge in the system came from a additional B or N atom replacing a C atom, which is not the case, then our calculations would be in agreement with Lieb's theorem for the half-mettallic states, but not the regular metals. This comes from the fact that, in half-metals, the additional electron or hole occupies only one kind of band (spin up or down), whereas in regular metals both bands are occupied, thus reducing the value of $M$.  Finally, since the spin density is concentrated in one of the clusters in the charged cases, we have a FM coupling between second nearest neighbor clusters and a FM coupling between neighboring supercells for these cases. For the neutral cases, however, we have an AF coupling between first neighbor clusters and no coupling between neighbor supercells (because $M = 0$).

\section{\label{conclusion}Conclusions}

In conclusion, we have studied the electronic and magnetic properties of h-BN triangular clusters embedded in graphene, with a particular supercell geometry, in which we have two clusters with opposite type of edges (B or N terminated) and neighboring C atoms occupying different sublattices. For this system, we have found 3 different types of band structure, depending on the bandwidth of the bands and their proximity to the neutral Fermi level. The energy gap is found to decrease with both increasing cluster and graphene separation region sizes, while the bandwidth of the closest bands depends only on the former, also decreasing with increasing cluster size. For intermediary and large cluster sizes, the band structure exhibits spin polarization in the charged states and half-metallic states are observed for certain charge values. These states are characterized by a strong spin density concentrated on one of the clusters, depending on the sign of the injected charge, and a strong net magnetization in the supercell, corresponding to the full value of the injected charge. This is related to the nature of the occupied and unnocupied bands near the neutral Fermi level, where the occupied (unoccupied) bands are localized levels formed mainly by a hybridization of $p_z$ orbitals from C and B (N) atoms near on the edges of each cluster. For large clusters in the neutral state, both set of bands show spin polarization and an AF coupling between neighboring clusters is observed, which resembles the spin density observed in intercalated h-BN/graphene zigzag ribbons calculations in the same size range. This system offers the possibility of band gap engineering through control of cluster and graphene region sizes, in a similar fashion to other calculations with quantum dots of graphene or dots embedded in graphene. We also have the possibility of controling its electric behavior (semiconductor, metallic or half-metallic) through control of both sizes and charge doping. We point out, however, that we have only studied a particular geometry for h-BN domains in graphene, given their triangular shape, their alignment and the intrinsic periodicity of the supercell. The domains already observed experimentally show no such properties, having a general shape with combinations of armchair and zigzag edges and mixed terminations (B or N bonding with C). Nevertheless, these domains are found to have band gap engineering properties similar to those observed here, which are a consequence of the mixing between the bulk h-BN and pure graphene properties. On the other hand, the finer details, such as the spin polarization trends, will be strongly dependent on the character of the different type of edges in a general domain. For example, zigzag edges related to carbon atoms in different sublattices (A or B) and with different terminations (B or N) tend to cancel their spin polarizations if they are too close to each other, as expected from their AF coupling. Edges of the same type tend to have FM coupling, but they may also cancel out if too close, probably due to rehybridization, such as in the vertices of triangular clusters. The number of localized states near the Fermi level will also depend on the shape of the cluster and will be related to the imbalance in number of C atoms missing from the A and B sublattices and their respective occupations by B or N atoms. The existence of half-metallic states and spin density properties will be related to these levels and we expect to see similar trends to those observed here in the case of zigzag edges. Therefore, we expect this work to offer a contribution in the understanding of the electronic and magnetic properties of h-BN domains in graphene, specially for the regions with zigzag edges, and to pave the way for future applications of such heterostructures in nanoelectronics and spintronics.

\begin{acknowledgments}
We thank Professor Steven G. Louie and his group for valuable discussions. We also acknowledge the financial support from the Brazilian funding agencies: CAPES, CNPq, FAPERJ and INCT-Nanomateriais de Carbono.
\end{acknowledgments}

\bibliography{Artigo}

\end{document}